\documentclass[draft]{agujournal2019}
\usepackage{url}
\usepackage{lineno}
\usepackage[inline]{trackchanges}
\usepackage{soul}
\draftfalse
\usepackage{xcolor}

\journalname{JGR: Space Physics}

\usepackage{amsmath,amssymb,amsfonts,amstext,graphics,graphicx,txfonts,wrapfig}

\def\bq{\begin{equation}}
\def\eq{\end{equation}}
\def\bqy{\begin{eqnarray}}
\def\eqy{\end{eqnarray}}
\def\p{\partial}

\definecolor{MyBrown}{cmyk}{.40,.65,.90,.35}
\definecolor{MyBlue}{cmyk}{.70,.15,.00,.00}

\def\BlueA{\setlength{\fboxsep}{0mm}\fcolorbox{black}{MyBlue}{\color{white} (a)}}
\def\BrownA{\setlength{\fboxsep}{0mm}\fcolorbox{black}{MyBrown}{\color{white} (a)}}

\def\BrownAtot{\setlength{\fboxsep}{0mm}\fcolorbox{black}{MyBrown}{\color{white} tot,(a)}}
\def\BlueBtot{\setlength{\fboxsep}{0mm}\fcolorbox{black}{MyBlue}{\color{white} tot,(b)}}

\def\Bluetot{\setlength{\fboxsep}{0.2mm}\fcolorbox{black}{MyBlue}{\color{white} tot}}
\def\Browntot{\setlength{\fboxsep}{0.2mm}\fcolorbox{black}{MyBrown}{\color{white} tot}}

\begin{document}

\title{Pure Interchange Oscillations of Thin Filaments in an Average Magnetosphere}
\authors{F. R. Toffoletto\affil{1}, R. A. Wolf\affil{1} and J. Derr\affil{1}}
\affiliation{1}{Physics and Astronomy Department, Rice University}
\correspondingauthor{J. Derr}{jderr@rice.edu}

\begin{keypoints}
	\item We extend the theory of classic interchange by calculating eigenmodes within a thin filament approximation.
	\item We compare the eigenmodes from interchange theory to those from ideal MHD in an average magnetosphere with zero ionospheric conductance.
	\item The eigenmodes, which differ in the plasma sheet and inner magnetosphere, agree roughly for interchange and MHD calculations.
\end{keypoints}

\begin{abstract}

This paper describes magnetospheric waves of very long wavelength in thin magnetic filaments. We consider an average magnetospheric configuration with zero ionospheric conductance and calculate waves using two different formulations: classic interchange theory and ideal MHD. Classic interchange theory, which is developed in detail in this paper, is basically analytic and is relatively straightforward to determine computationally, but it can’t offer very high accuracy.

The two formalisms agree well for the plasma sheet and also for the inner magnetosphere. The eigenfrequencies range over about a factor of seven, but the formulations generally agree with a root-mean-square difference between the logarithms of interchange and MHD frequencies to be $\sim 0.054$. The pressure perturbations in the classic interchange theory are assumed constant along each field line, but the MHD computed pressure perturbations along the field line vary in a range $\sim 30 \%$ in the plasma sheet but are larger in the inner magnetosphere. The parallel and perpendicular displacements, which are very different in the plasma sheet and inner magnetosphere, show good qualitative agreement between the two approaches. In the plasma sheet, the perpendicular displacements are strongly concentrated in the equatorial plane, whereas the parallel displacements are spread through most of the plasma sheet away from the equatorial plane; and can be regarded as buoyancy waves. In the inner magnetosphere, the displacements are more sinusoidal and are more like conventional slow modes. The different forms of the waves are best characterized by the flux tube entropy $PV^\gamma$.

\end{abstract}

\section*{Plain Language Summary}
Ideal magnetohydrodynamic (MHD) ballooning and interchange disturbances have been studied extensively over the years as they are connected to potentially important phenomena in the magnetosphere. One such phenomenon consists of magnetospheric buoyancy waves which are analogous to atmospheric gravity waves, with magnetic tension force replacing gravity as the restoring force. There are different definitions of ideal ballooning and interchange. In our definition, ballooning is much more general, while interchange applies only to a much more limited set of conditions, as it applies to the limit of long wavelengths and assumes pressure is constant along each field line. Focusing on waves and using a thin filament approach, we have extended the classic theory of interchange by calculating eigenfrequencies and eigenfunctions for pure interchange models using an energy approach. The results are applied on an average force-balanced magnetosphere configuration and are compared to MHD ballooning oscillation eigenfrequencies and eigenfunctions. The two approaches show good agreement in the plasma sheet, but less so in the inner magnetosphere, where the MHD results qualitatively resemble MHD classic slow modes rather than buoyancy waves.

\section{Introduction}

This paper explores the relationship between two kinds of linear MHD disturbances, namely interchange and ideal MHD ballooning. We consider the limit of zero ionospheric conductance. In that case, the square of the frequency $\omega^2$ is real. If $\omega^2 > 0$, it represents a wave, and $\omega^2 < 0$, it represents a instability. In this paper, we are mostly concerned with waves. For simplicity, we consider even waves with $k_\parallel \sim 0$ propagating in the meridional $xz$ plane. (An even wave is one in which the displacement is symmetric about the equatorial plane.) We define $k_\parallel \coloneqq 2 \pi / \lambda$, where $\lambda$ is the wavelength of the thin filament mode.

We consider waves in thin magnetic filaments, which have been studied in various geometries (e.g., solar \cite{PARKER1981}; and magnetosphere \cite{CHEN1999, WOLF2012, TOFFO2020}). These structures are infinitesimally thin in the $y$-direction and also in the $xz$ plane perpendicular to the magnetic field. They slide through the magnetospheric background without friction. In a uniform medium, there are three MHD wave modes (fast, intermediate, and slow). The fast modes don't propagate in thin filaments \cite{CHEN1999}. In the intermediate (Alfv{\'e}n) modes, which have been studied exhaustively in the magnetosphere, the perturbation velocities are in the $\pm y$ direction in the $xz$ plane. We are most interested in the slow modes with perpendicular velocities in the $xz$ plane, particularly the ones with very long wavelengths along the magnetic field ($k_\parallel \approx 0$). In the $xz$ plane, the Alfv{\'e}n and slow modes decouple, and we discuss just the slow modes.

Many space physicists have discussed ballooning and interchange (e.g. \citeA{PANOV2022,KHAZANOV2020,SORATHIA2020,SOUTHWOOD1989,MAZUR2013,LIU1997,COWLEY2015,SCHINDLER2004,PRITCHETT2010,BIRN2011}). In our definition, ballooning is a much more general phenomenon than interchange. It includes both short and long wavelengths. Interchange applies only to a much more limited set of conditions. Specifically, it applies to the limit of long wavelengths  ($k_\parallel \approx 0$), and it also assumes pressure is constant along each field line. The theory of ballooning does not make that assumption. The theory of interchange is much easier to apply than the theory of ballooning. Interchange theory was developed more than sixty years ago by \citeA{BERNSTEIN1958}, in a classic paper written early the controlled-fusion effort. \citeA{BERNSTEIN1958} included a full range of $\beta$. A much simpler version of interchange theory was based on the energy principle, a version that assumes low-$\beta$, was developed \citeA{CHANDRA1960} and used by many textbook authors (e.g., \citeA{SCHMIDT1979}). Most of this early work emphasized the threshold of interchange instability. Our present paper, which is also based on the energy principle, uses the thin-filament approximation for realistic magnetospheres. It is not limited to calculating the threshold $\omega^2 = 0$, and it includes the calculation of the eigenfrequencies and eigenfunctions.

In a neutral atmosphere, the buoyancy force is due to gravity and is proportional to the entropy $P / \rho^\gamma$. If the gradient of $P / \rho^\gamma$ is downward, the system is interchange unstable. If the gradient of $P / \rho^\gamma$ is upward, the system can show stable interchange waves. In the magnetosphere, gravity is unimportant, but there is an effective buoyancy that is proportional to the curvature of the magnetic field lines. For this magnetic buoyancy, the crucial physical quantity is the flux tube entropy $P V^\gamma$. If the gradient of $P V^\gamma$ is earthward the system is unstable to interchange. If the gradient of $P V^\gamma$ is anti-earthward, the result is stable waves. In the simplest case, the equation that governs the frequency of the oscillation is the same as the Taylor-Goldstein equation, which governs the oscillations of the neutral atmosphere. See \citeA{WOLF2018} for details.

In this paper, we consider wave motion as being confined to the $xz$ plane. The case in which the gradient is not on a plane has been considered theoretically \cite{XING2007, KHAZANOV2020} but the results conflict, and the situation has not been resolved. Section 2 of the paper uses analytic theory and the energy principle to derive formulas for the eigenfunctions and eigenfrequencies within the classic interchange theory. Section 3 displays the results of numerical calculations of eigenfunctions and eigenfrequencies, comparing the classic interchange theory with full MHD calculations for an average magnetosphere. Section 4 discusses the remarkable differences between plasma sheet and inner magnetosphere, for an average magnetosphere and the relationship between the flux tube entropy $P V^\gamma$, the boundary between the unstable and stable region, and the accuracy by which we can estimate eigenfrequencies and eigenfunctions using full MHD and the interchange approximation. An appendix explains a numerical procedure to determine inner boundary conditions for the case of zero ionospheric conductance.

\section{Formulas for the Amplitudes and Buoyancy Frequency for Pure Interchange Oscillation of Thin Filaments}

Here we use an energy argument to derive the expressions for the field-transverse and field-aligned displacement eigenfunctions ($\xi_\perp$, $\xi_\parallel$) and eigenfrequency $\omega_{PI}$ for the pure interchange oscillations of a thin filament. The interchange picture is idealized in the sense that it assumes that $\xi_\perp(s)$ displaces an equilibrium field line to the shape of an adjacent equilibrium field line and also that $\xi_\parallel$ maintains pressure constancy along the displaced flux tube. These can be compared with the eigenfunctions and eigenfrequencies derived for MHD normal modes of the thin filament to determine how closely those normal modes resemble pure interchange modes, which are one type of idealization of a thin filament normal mode.

\begin{figure}[h!]
	\centering
	\textbf{Two Thin Filaments Set Oscillating by Interchange}
	\includegraphics[width=1\textwidth]{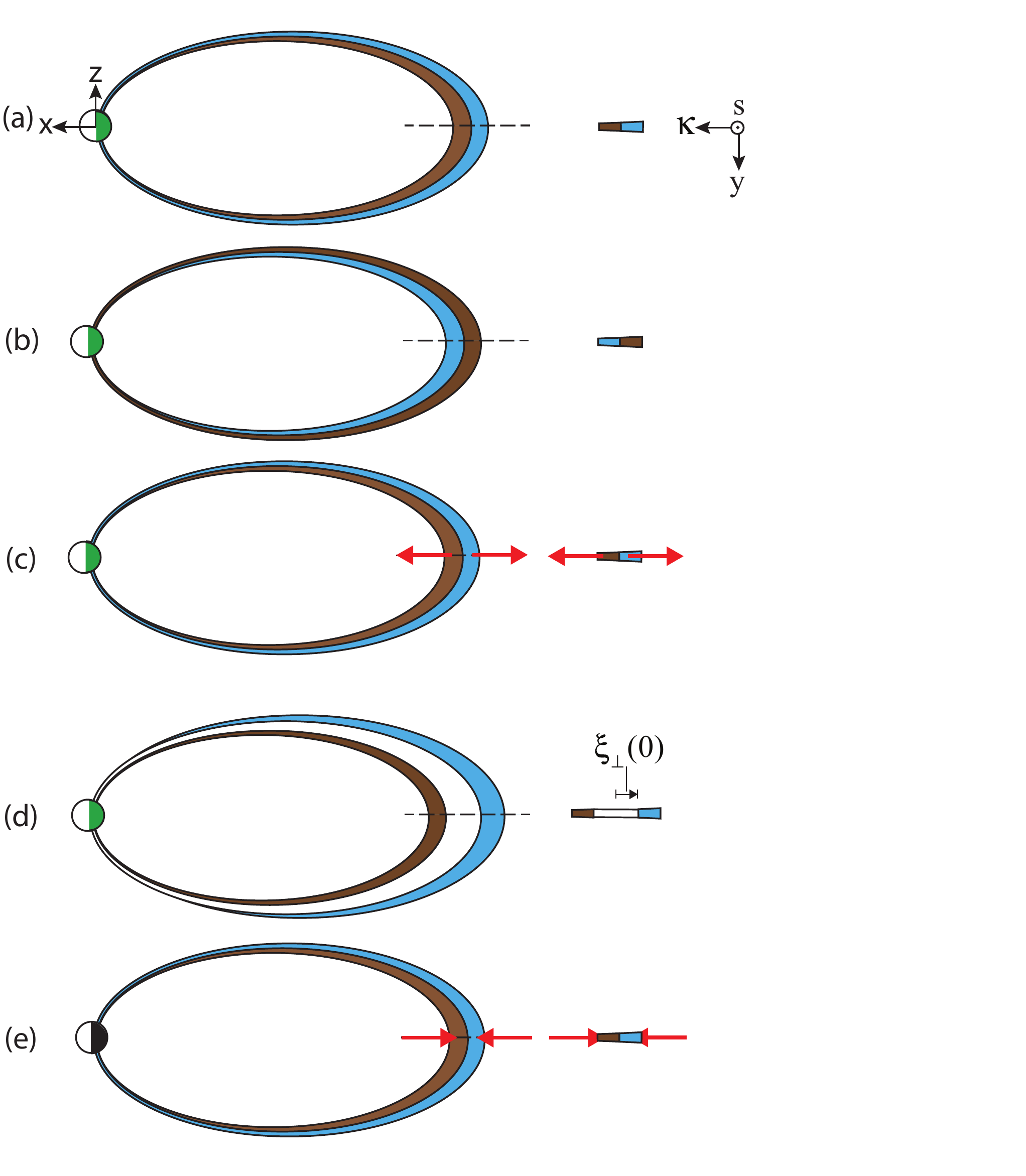}
	\caption{Meridional (left) and equatorial (right) views of the oscillating 2-filament problem we are considering. The two filaments (brown and blue) contain identical amounts of magnetic flux $\Phi$ and have the same flux tube entropy $P V^{5/3}$ as the local background. The entropy gradients are anti-Earthward (to the right in both images).}
	\label{Figure1}
\end{figure}

Consider the idealized situation of two adjacent thin filaments shown in Figure \ref{Figure1}. In diagram (a), we start with two thin filaments from the background that have the same magnetic flux $\Phi$. Each matches the local background, so the system is in equilibrium. Between diagrams (a) and (b), an external force pushes the blue filament past the brown filament, displacing it to the flux tube volume previously occupied by the brown filament. The filaments have switched locations and the external force withdraws, leaving each filament occupying the same volume as the other had initially. Neither one is left with kinetic energy. However, they are now out of equilibrium with their surroundings. The gradient of $P V^\gamma$ propels the blue filament anti-Earthward and the brown filament Earthward. When they reach the positions they had originally occupied, they now have residual kinetic energy which propels them from their locations in (c) to their locations in (d). By configuration (d), they have converted all of their kinetic energy into magnetic energy and have come to rest. This energy now propels them in the opposite directions, blue Earthward and brown anti-Earthward. When they reach configuration (e), the buoyancy forces have been reduced to zero and the velocities of the equatorial footprints have reversed relative to (c). These oscillations continue to repeat in the absence of any additional disturbances or damping.

We will here analyze the oscillations of the thin filaments from an energy point of view under the interchange assumption. In configuration (b), the total energy is entirely potential, whereas, in configuration (c), it is entirely kinetic. Equating the potential energy in configuration (b) to the kinetic energy in configuration (c) will reveal the frequency of the oscillation.

First, let us compute the potential energy of configuration (b).

\subsection{Potential Energy}

We calculate the change in potential energy in two steps. In the first step, we assume the blue filament in configuration (b) has moved to occupy the same volume as the brown filament had in configuration (a), and vice-versa. Since the two filaments have the same magnetic flux, the total magnetic energy is unchanged from (a) to (b).

\subsubsection{Step 1: Potential Energy Change Due to Exchange of Thin Filaments}

We calculate the change in potential energy going from configuration (a) to configuration (b). Configurations (a) and (c) are identical with respect to their kinetic energy, so we will use the (a) configuration label from now on rather than (c). The particle pressure part of the internal energy of the two filaments (distinguished by their color box subscripts), in their initial configuration (a), is given by:

\bq
	U_{(a)} = U_{(c)} = \frac{3}{2} \left( P_{\BrownA} V_{\BrownA} + P_{\BlueA} V_{\BlueA} \right) \Phi.
\eq

where $\Phi$ is the amount of magnetic flux in each flux tube. If we assume the process is adiabatic, the particle pressure part of the internal energy of the two filaments in configuration (b) (in terms of the pressures and flux tube volumes which they occupied in configuration (a)) is given by:

\bq
	U_{(b)} = \frac{3}{2} \left[ P_{\BlueA} \left( \frac{V_{\BlueA}}{V_{\BrownA}} \right)^{5/3} V_{\BrownA} + P_{\BrownA} \left( \frac{V_{\BrownA}}{V_{\BlueA}} \right)^{5/3} V_{\BlueA} \right] \Phi.
\eq

To first-order, the difference between the initial pressures and flux tube volumes in the brown and blue filaments is given by:

\bqy
	V_{\BlueA} &=& V_{\BrownA} + V' \xi_{\perp} \\
	P_{\BlueA} &=& P_{\BrownA}+ P' \xi_{\perp},
\eqy

where the primes represent the derivatives in the tailward/poleward direction. The resulting change in potential energy to second order in the small parameter $\xi$ can therefore be written:

\bq
	U_{(b)} - U_{(a)} = P V \Phi \xi_{\perp}^2 \frac{V'}{V} \left( \frac{P'}{P} + \frac{5}{3} \frac{V'}{V} \right). \label{5}
\eq

\subsubsection{Step 2: Pressure Equilibrium Restoration by Adjacent Thin Filament Boundary Adjustment}

At this point, in configuration (b), the total pressure inside each of the filaments does not balance the local background pressure, because while the magnetic field inside the blue and brown filaments has not yet changed in the interchange, the particle pressure has. This is not important in very low $\beta$ plasmas, but it is in high $\beta$ plasmas.

Without loss of generality, let us consider the blue filament closer to the Earth in configuration (b). We will find the additional pressure available to this filament as a result of the exchange.

The total pressure in the filament closer to Earth in configurations (a) and (b), respectively, are given by:

\bqy
	P_{\BrownAtot} &=& \frac{B_{\BrownA}^2}{2 \mu_0} + P_{\BrownA} \\
	P_{\BlueBtot} &=& \frac{B_{\BrownA}^2}{2 \mu_0} + P_{\BlueA} \left( \frac{V_{\BlueA}}{V_{\BrownA}} \right)^{5/3},
\eqy

so that to first order (dropping the color box subscripts), we have:

\bq
	\delta P_{\Bluetot} \coloneqq P_{\BlueBtot} - P_{\BrownAtot} = P \xi_\perp \left( \frac{P'}{P} + \frac{5}{3} \frac{V'}{V} \right). \label{8}
\eq

Since the gradient of $PV^{5/3}$ is anti-Earthwards, it follows that $\delta P_{\Bluetot} > \delta P_{\Browntot}$. The blue filament is no longer in equilibrium with its surroundings, so it will have to expand. The brown filament will have to contract by an equal amount.

\begin{figure}[h!]
	\centering
	\textbf{Pressure Re-equilibration by Expansion and Contraction of Thin Filaments}
	\includegraphics[width=1\textwidth]{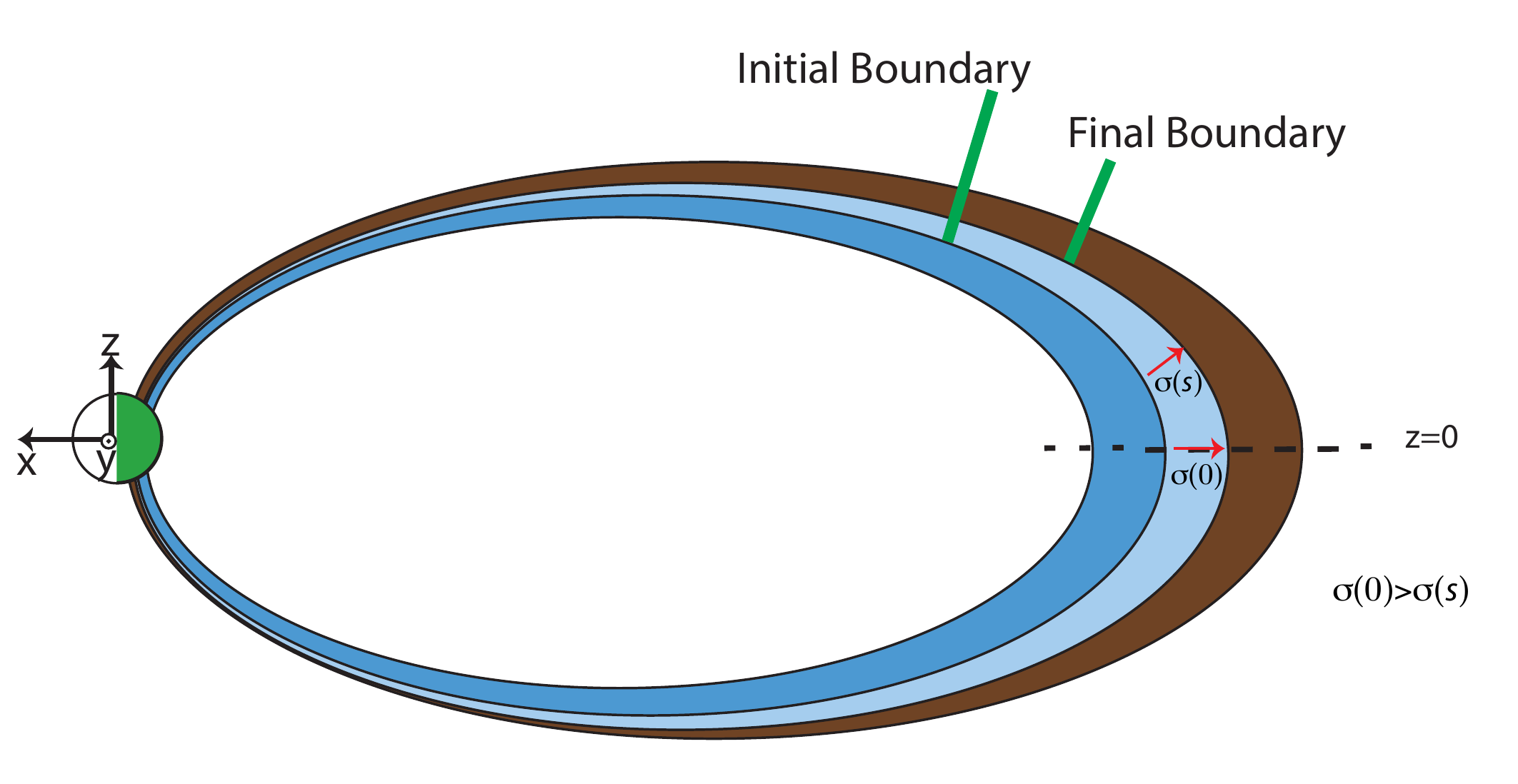}
	\caption{Meridional diagram of two adjacent thin filaments after they have exchanged locations. The thin filament closer to the Earth is not in equilibrium and will have to expand to reach equilibrium as the other filament contracts. This shift of the boundary $\sigma(s)$ between the two thin filaments depends on the field-aligned coordinate $s$ and takes its maximum value $\sigma(0)$ in the equatorial plane. The entire initial and final boundary location is shown. In the equatorial plane, the boundary has shifted anti-Earthward corresponding to the poleward boundary shift of the footprint on the Earth. Recall that the ionospheric conductance is assumed to be zero.}
	\label{Figure2}
\end{figure}

To simplify things, we do not consider the azimuthal motion of the filaments. They are infinitesimally thin azimuthally and infinitesimal motion in this direction introduces no changes in potential energy. Therefore the figures do not include the real, if unimportant, azimuthal sliding of the filaments around one another so that they may exchange locations.

With the expansion of the blue filament towards equilibrium, the boundary between the brown and blue filaments needs to move up in the diagram. However, the boundary between the two filaments extends along the entire length of the filament, and the boundary location adjustment $\sigma$ depends on the distance along the field line from the equatorial plane, which we designate $s$.

Again, consider the blue filament in configuration (b) (with superfluous subscripts dropped). We allow the boundary between the two filaments to shift by $\sigma(s)$ toward the brown filament as seen in Figure \ref{Figure2}, so that:

\bq
	B_{f}(s) \left[ \xi_\perp(s) + \sigma(s) \right] = B \, \xi_\perp(s),
\eq

where $\xi_{\perp}(s)$ is tailward and positive and $B_{f}$ is the final value of the magnetic field in the blue filament in configuration (b) after the pressure adjustment. The fractional perturbation of the magnetic inductance is thus given by:

\bq
	\frac{\delta B}{B} = - \frac{\sigma(s)}{\xi_\perp(s)},
\eq

where we continue to drop higher order contributions. The change in magnetic pressure of the blue filament, due to the boundary adjustment is, to first order,

\bq
	\delta \left( \frac{B^2}{2 \mu_0} \right) = - \frac{B^2}{\mu_0} \frac{\sigma(s)}{\xi_\perp(s)}.
\eq

The corresponding fractional change in flux tube volume is:

\bq
	\frac{\delta V}{V} = - \left< \frac{\delta B(s)}{B} \right>_{ft} = \left< \frac{\sigma(s)}{\xi_\perp(s)} \right>_{ft},
\eq

where the angle bracket represents a flux tube average:

\bq
	\left< f \right>_{ft} = \frac{1}{V} \int \frac{ds}{B} f(s).
\eq

Since we assume adiabaticity, we arrive at the following fractional change in particle pressure:

\bq
	\frac{\delta P}{P} = - \frac{5}{3} \left< \frac{\sigma}{\xi_\perp} \right>_{ft}.
\eq

Finally, the fractional change in the total pressure due to the boundary adjustment is:

\bq
	\frac{\delta P_{tot}(s)}{P} = - \frac{5}{3} \left< \frac{\sigma}{\xi_\perp} \right>_{ft} - \frac{B^2}{\mu_0 P} \frac{\sigma(s)}{\xi_\perp(s)}. \label{15}
\eq

It would be advantageous to recast this exclusively in terms of the equatorial plane values rather than as a general function of $s$.

This pressure adjustment just needs to balance out the additional pressure given by (\ref{8}), which means that

\bq
	\frac{B^2}{\mu_0 P} \frac{\sigma(s)}{\xi_\perp(s)} + \frac{5}{3} \left< \frac{\sigma}{\xi_\perp} \right>_{ft} = \xi_\perp(s) \left[ \frac{P'(s)}{P} + \frac{5}{3} \frac{V'(s)}{V} \right]. \label{16}
\eq

Since $P$ and $V$ are both constant along field lines and $\xi_\perp$ is proportional to the distance between nearby field lines, the right side of (\ref{16}) is constant along field lines, and we can evaluate it in the equatorial plane (designated by an ``e'' subscript). Doing this and solving for the fractional shift in flux tube boundary, we arrive at the following expression:

\bq
	\frac{\sigma(s)}{\xi_\perp(s)} = \beta(s) \left( - \frac{5}{6} \left< \frac{\sigma}{\xi_\perp} \right>_{ft} + \frac{\xi_\perp(0)}{2} \left[ \frac{P_e'}{P} + \frac{5}{3} \frac{V_e'}{V} \right] \right). \label{17}
\eq

Taking the flux tube average of both sides gives and solving for the flux tube averaged fractional boundary shift, we obtain:

\bq
	\left< \frac{\sigma}{\xi_\perp} \right>_{ft} = \frac{\xi_\perp(0)}{2} \frac{\left< \beta \right>_{ft}}{\left( 1 + \frac{5}{6} \left< \beta \right>_{ft} \right)} \left( \frac{P_e'}{P} + \frac{5}{3} \frac{V_e'}{V} \right). \label{18}
\eq

Now, let's substitute (\ref{18}) into (\ref{17}) to arrive at:

\bq
	\frac{\sigma(s)}{\xi_\perp(s)} = \frac{\xi_\perp(0)}{2} \frac{\beta(s)}{\left( 1 + \frac{5}{6} \left< \beta \right>_{ft} \right)} \left( \frac{P_e'}{P} + \frac{5}{3} \frac{V_e'}{V} \right). \label{19}
\eq

Substitution of (\ref{18}) and (\ref{19}) into (\ref{15}) finally yields an expression for the fractional change in total pressure which depends only on parameters evaluated at the equatorial plane:

\bq
	\frac{\delta P_{tot}}{P} = - \xi_\perp(0) \left( \frac{P_e'}{P} + \frac{5}{3} \frac{V_e'}{V} \right).
\eq

Now that we know how much the boundary between the blue and brown filaments moves to bring the total pressures into equilibrium, we need to know how much that boundary shift reduces the potential energy. Imagine there is a horizontal wall between the two filaments in Figure \ref{Figure2}. The work done on this wall would be:

\bq
	\delta W = - \frac{1}{2} \int ds \, \Delta y(s) \, \sigma(s) \left( \delta P_{\Bluetot} - \delta P_{\Browntot} \right) = - \int ds \, \Delta y(s) \, \sigma(s) \, \delta P_{\Bluetot}, \label{21}
\eq

where $\Delta y(s)$ is the width of the filament in the $y$-direction, which is perpendicular to the plane of the field lines. The factor of $1/2$ comes from the fact that the pressure imbalance goes from its maximum value of zero as the wall moves to its equilibrium position, and we have in the last step used the fact that $\delta P_{\Browntot} = - \delta P_{\Bluetot}$. Now that we have eliminated the brown color box subscript, we will again dispense with the blue color box subscript, which should be now again be taken as implicit. Now, we can utilize the fact that $\Delta y(s) \, \xi_{\perp}(s) = \Phi/B(s)$, so that (\ref{21}) becomes:

\bq
	\delta W = - \Phi \int \frac{ds}{B} \frac{\sigma(s)}{\xi_\perp(s)} \, \delta P_{tot} = - \frac{1}{2} P V \Phi \, \xi_\perp^2(0) \frac{\left< \beta \right>_{ft}}{\left( 1 + \frac{5}{6} \left< \beta \right>_{ft}  \right)} \left( \frac{P_e'}{P} + \frac{5}{3} \frac{V_e'}{V} \right)^2. \label{22}
\eq

The total difference between the potential energy in state (b), with the boundary adjustment included, and state (a) is:

\bq
	\Delta U = U_{(b)} - U_{(a)} + \delta W.
\eq

Using (\ref{5}) and (\ref{22}), we therefore obtain our final expression for the change in potential energy:

\bq
	\Delta U = P V \Phi \, \xi_\perp(0)^2 \frac{1}{1 + \frac{5}{6} \left< \beta \right>_{ft}} \left( \frac{P_e'}{P} + \frac{5}{3} \frac{V_e'}{V} \right) \left( \frac{V_e'}{V} - \frac{\left< \beta \right>_{ft}}{2} \frac{P_e'}{P} \right). \label{24}
\eq

The last two factors determine the sign of $\omega^2$. They are consistent with equation (2) of \citeA{XING2007}, which came from equation (6.45) of \citeA{BERNSTEIN1958}.

\subsection{Kinetic Energy}

Now that we have computed the total change in potential energy, let us turn to the maximum kinetic energy of the filaments, which they have at configurations (c) and (e), for example, of Figure \ref{Figure1}.

\begin{figure}[h!]
	\centering
	\textbf{Noon-Midnight Meridional Geometry of Adjacent Thin Filaments} \\
	\textbf{} \\
	\includegraphics[width=0.7\textwidth]{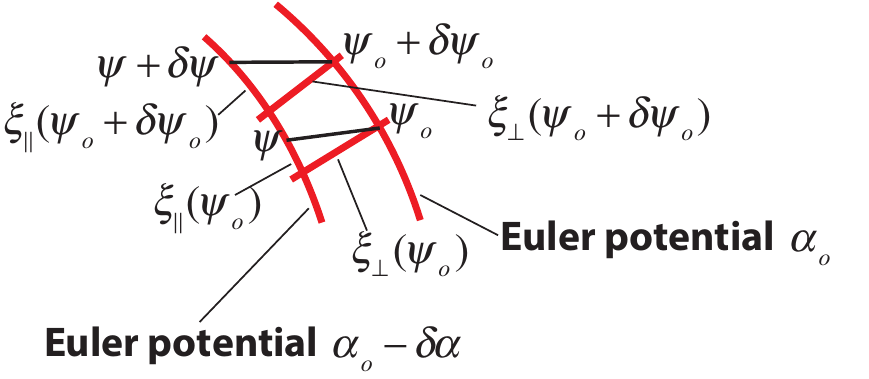}
	\caption{Geometry of adjacent thin filaments in Euler potential $\alpha$ used in the calculation of $\xi_\parallel$. We imagine this to be the noon-midnight meridional plane, so that the $y$-coordinate is fixed. Coordinate grid-lines are drawn in the $(\alpha,\psi)$-plane.}
	\label{Figure3}
\end{figure}

For an interchange motion, the displacement perpendicular to the field line is given by:

\bq
	\xi_\perp = - \xi_\kappa = \frac{\delta \alpha}{|\nabla \alpha|(s)} \equiv h_\alpha \delta \alpha
\eq

where $\alpha$ is the Euler potential that specifies the field line. This equation guarantees that if the initial filament location coincides with a background field line that is labeled $\alpha = \alpha_0$, then the displaced filament coincides with a background field line that is labeled $\alpha = \alpha_0 - \delta \alpha$. We apply this to the midnight meridian ($y = 0$). The $s$-dependence of $\xi_\parallel$ is determined by the assumption that pressure is constant on the field line defined by the displaced fluid elements.

The geometry is shown in Figure \ref{Figure3}. We use a second coordinate $\psi$, which is used to measure the distance along the field line. Its gradient is in the midnight meridian plane and is orthogonal to $\nabla \alpha$. The physical distance between $\psi_0$ and $\psi_0 + \delta \psi_0$ on the field line $\alpha$ is $h_\psi(\alpha_0, \psi_0) \, \delta \psi_0$ and the physical distance along a $\psi = $ constant line between field line $\alpha$ and field line $\alpha - \delta \alpha$ is $\xi_\perp(\alpha, \psi) = h_\alpha(\alpha, \psi) \, \delta \alpha$.

The number of particles on field line $\alpha_0$ between $\psi_0$ and $\psi_0 + \delta \psi_0$ has to be equal to the number of particles on field line $\alpha_0 - \delta \alpha$ between $\psi$ and $\psi + \delta \psi$, so that:

\bq
	\frac{n(\alpha_0 - \delta \alpha)}{B(\alpha_0 - \delta \alpha,\psi)} \, h_\psi(\alpha_0 - \delta \alpha_,\psi) \, \delta \psi = \frac{n(\alpha_0)}{B(\alpha_0,\psi_0)} \, h_\psi(\alpha_0,\psi_0) \, \delta \psi_0.
\eq

From Figure \ref{Figure3}, we can see that the relationship between $\delta \psi$ and $\delta \psi_0$ is given by:

\bq
	h_\psi(\alpha_0 - \delta \alpha,\psi) \, \delta \psi = h_\psi(\alpha_0,\psi_0) \, \delta \psi_0 \left[ 1 - \kappa h_\alpha \delta \alpha + \frac{1}{h_\psi} \left( \frac{\p \xi_\parallel}{\p \psi} \right)_\alpha \right],
\eq

where $\kappa$ is the field line curvature. Flux tube particle conservation, with the application of this geometric relation, can be written:

\bq
	\frac{n(\alpha_0 - \delta \alpha)}{B(\alpha_0 - \delta \alpha, \psi_0 + \delta \psi)} \left( 1 - \kappa h_\alpha \delta \alpha + \frac{1}{h_\psi} \left( \frac{\p \xi_\parallel}{\p \psi} \right)_\alpha  \right) = \frac{n(\alpha_0)}{B(\alpha_0,\psi_0)}. \label{28}
\eq

The difference between the magnetic fields that a given fluid element will experience in going from $(\alpha_0, \psi_0)$ to $(\alpha_0 - \delta \alpha, \psi_0 + \delta \psi)$ is:

\bq
	\frac{B(\alpha_0 - \delta \alpha,\psi_0 + \delta \psi)}{B(\alpha_0, \psi_0)} = 1 - \frac{1}{B} \left( \frac{\p B}{\p \alpha} \right)_\psi \delta \alpha + \frac{1}{B} \left( \frac{\p B}{\p \psi} \right)_\alpha \left( \delta \psi - \delta \psi_0 \right). \label{29}
\eq

Conservation of particles in the context of frozen-in flux require that:

\bq
	\frac{n(\alpha_0 - \delta \alpha)}{n(\alpha_0)} = \frac{V(\alpha_0)}{V(\alpha_0 - \delta \alpha)} \approx 1 + \frac{1}{V} \frac{dV}{d\alpha} \delta \alpha. \label{30}
\eq

We use (\ref{29}) to eliminate magnetic inductance, (\ref{30}) to eliminate the densities, and the fact that $\delta \psi - \delta \psi_0 = \xi_\parallel / h_\psi$ to rewrite (\ref{28}) in the following form, retaining only first-order terms in $\delta \alpha$:

\bq
	\frac{1}{h_\psi} \left[ \left( \frac{\p \xi_\parallel}{\p \psi} \right)_\alpha - \frac{\xi_\parallel}{B} \left( \frac{\p B}{\p \psi} \right)_\alpha \right] = \frac{1}{h_\psi} \left( \frac{\p \left( \xi_\parallel / B \right)}{\p \psi} \right)_\alpha = \left[ \kappa h_\alpha - \frac{1}{V} \frac{dV}{d\alpha} - \frac{1}{B} \left( \frac{\p B}{\p \alpha} \right)_\psi \right] \delta \alpha.
\eq

This can easily be integrated in $\psi$ to yield:

\bq
	\frac{\xi_\parallel(\psi)}{B(\psi)} = \delta \alpha \int_0^\psi \frac{h_\psi(\psi') d\psi'}{B(\psi')} \left[ \kappa h_\alpha - \frac{1}{V} \frac{dV}{d\alpha} - \frac{1}{B} \left( \frac{\p B}{\p \alpha} \right)_{\psi'} \right], \label{32}
\eq

where we have made use of the fact that $\xi_\parallel(0) = 0$.

Now, we will make use of some geometry to recast our parallel displacement. We have:

\bq
	\left( \frac{\p h_\psi}{\p \alpha} \right)_\psi = \kappa h_\psi h_\alpha,
\eq

and we additionally note that one can commute the partial derivative past the integral (under an assumption of ``smoothness'') to write:

\bq
	\left( \frac{\p}{\p \alpha} \int_0^\psi \frac{h_\psi(\alpha,\psi') d\psi'}{B(\alpha,\psi')} \right)_\psi = \int_0^\psi \left( \frac{\p h_\psi(\alpha,\psi')}{\p \alpha} \right)_\psi \frac{d\psi'}{B} - \int_0^\psi \frac{h_\psi(\alpha,\psi') d\psi'}{B} \frac{1}{B} \left( \frac{\p B}{\p \alpha} \right)_{\psi'}
\eq

Technically, we are here assuming continuity of both $h_\psi / B$ and $\left[ \p_\alpha \left( h_\psi / B \right) \right]_\psi$ in the domain of interest.

Using these geometric constraints, we can rewrite equation (\ref{32}) as:

\bq
	\xi_\parallel(\psi) = B(\psi) \left[ \frac{\p V_p(\psi)}{\p \alpha} - \frac{V_p(\psi) }{V(\alpha)} \frac{dV(\alpha)}{d\alpha} \right] \delta \alpha, \label{35}
\eq

where:

\bq
	V_p(\alpha,\psi) = \int_0^\psi \frac{h_\psi(\alpha,\psi') d\psi'}{B(\alpha,\psi')}
\eq

is the partial flux tube volume from the equatorial plane up to coordinate $\psi$. Equation (\ref{35}) can now be rewritten in a slightly more general manner:

\bq
	\xi_\parallel(\psi) = B(\psi) \left[ \left( \frac{\p V_p(\alpha,\psi)}{\p s_\kappa} \right)_\psi - \frac{V_p(\psi) }{V(\alpha)} \left( \frac{\p V(\alpha)}{\p s_\kappa} \right)_\psi \right] \delta s_\kappa \label{37}
\eq

where $s_\kappa$ is the distance perpendicular to the field line in the direction of the curvature vector.

The maximum kinetic energy in the two flux tubes that each contain unit flux $\Phi$ can be written:

\bq
	T = \rho \omega_{PI}^2 \Phi \int \frac{ds}{B} \xi_\perp^2 \left( 1 + \frac{\xi_\parallel^2}{\xi_\perp^2} \right), \label{38}
\eq

where $\omega_{PI}$ is the pure interchange frequency associated with the thin filament oscillations.

\subsection{Eigenfrequency Formula}

Equating the maximum kinetic energy (\ref{38}) to the maximum potential energy (\ref{24}) and solving for the frequency gives:

\bq
	\omega_{PI}^2 = \frac{P}{\rho} \frac{ \left( \frac{P_e'}{P} + \frac{5}{3} \frac{V_e'}{V} \right) \left( \frac{V_e'}{V} - \frac{\left< \beta \right>_{ft}}{2} \frac{P_e'}{P} \right)}{\left( 1 + \frac{5}{6} \left< \beta \right>_{ft} \right) \left< \frac{\xi_\perp(s)^2}{\xi_\perp(0)^2} \left( 1 + \frac{\xi_\parallel(s)^2}{\xi_\perp(s)^2} \right) \right>_{ft}} \label{39}
\eq

where $s = 0$ represents the equatorial plane and $\rho = m n$ is the mass density of the thin filament plasma. This is the oscillation frequency for the pure interchange modes.

\section{Results}

\subsection{Average Magnetosphere}

\begin{figure}[h!]
	\centering
	\textbf{Equilibrium Magnetic Field Lines for Average Magnetosphere} \\
	\includegraphics[width=1\textwidth]{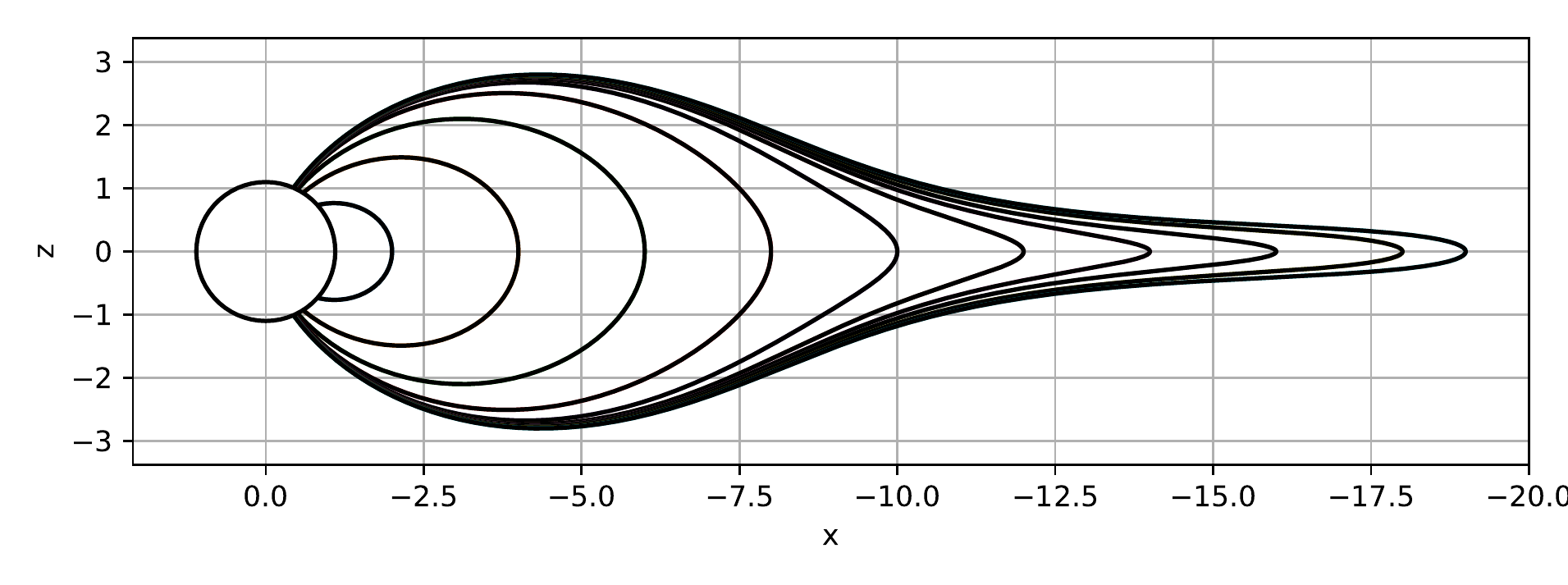}
	\caption{Plot of equilibrium field lines in the noon-midnight plane.}
	\label{Figure4}
\end{figure}

\begin{figure}[h!]
	\centering
	\textbf{Background Field Profiles for Average Magnetosphere} \\
	\includegraphics[width=1\textwidth]{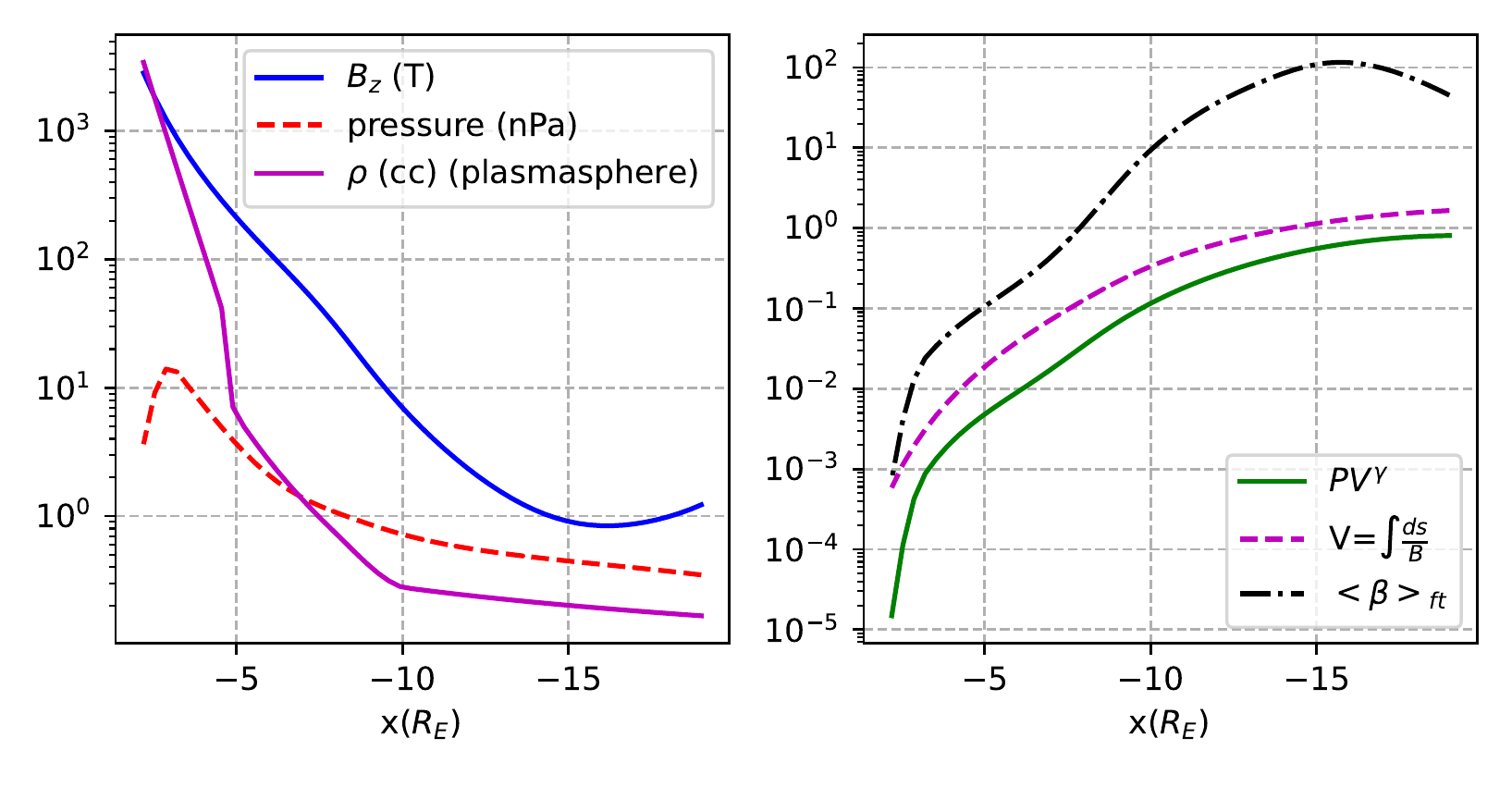}
	\caption{Plot of equilibrated magnetosphere models, showing on the left panel the pressure, $B_z$, and density (left) along the tail axis and the flux tube volume (V, right), entropy $PV^\gamma$ and the flux tube-averaged plasma beta $\left< \beta \right>_{ft}$.}
	\label{Figure5}
\end{figure}

We now plot the frequency $\omega_{PI}$ and several others, for a magnetosphere model which represents the $xz$ plane of an average magnetosphere. We use the following background fields:

\begin{enumerate}
	\item Started with a $Kp = 2$ \citeA{TSY1989} magnetic field model.
	\item Obtained pressure profile by combining a quiet curve from \citeA{LUI1987} for $|x| < 8 \ R_E$ and \citeA{SPENCE1989} for $|x| > 8 \ R_E$.
	\item Relaxed the resulting magnetic field and pressure configuration to equilibrium in the $xz$ plane using a 2-D, high-resolution version of the friction code \cite{LEMON2003}.
	\item Densities were chosen using the $Kp = 2$ \citeA{GALLAGHER2000} model for $|x| < 8 \ R_E$ merged smoothly to a \citeA{TSY2003} model for $|x| > 10 \ R_E$.
	\item For a specified equatorial crossing point, a field line is traced to both the Northern and Southern ionosphere.
\end{enumerate}

Figure \ref{Figure4}, reproduced from \citeA{TOFFO2020}, shows the corresponding field lines for the nightside of the background equilibrium. More details on the average magnetosphere model can be found in \citeA{WOLF2018}. Figure \ref{Figure5}, reproduced from \citeA{WOLF2018}, shows the resulting equilibrated magnetosphere from the equilibrium code of pressure, number density, and $B_z$ profiles along the tail axis, along with the flux tube volume, entropy $PV^\gamma$, and flux tube-averaged plasma beta $\left< \beta \right>_{ft}$. The magnetic field and pressure are also smoothed to remove any small grid-scale fluctuations introduced by the equilibrium code.

\subsection{Eigenfrequencies for an Average Magnetosphere}

\begin{figure}[h!]
	\centering
	\textbf{Buoyancy Frequencies as a Function of Equatorial Crossing Distance}
	\includegraphics[width=1\textwidth]{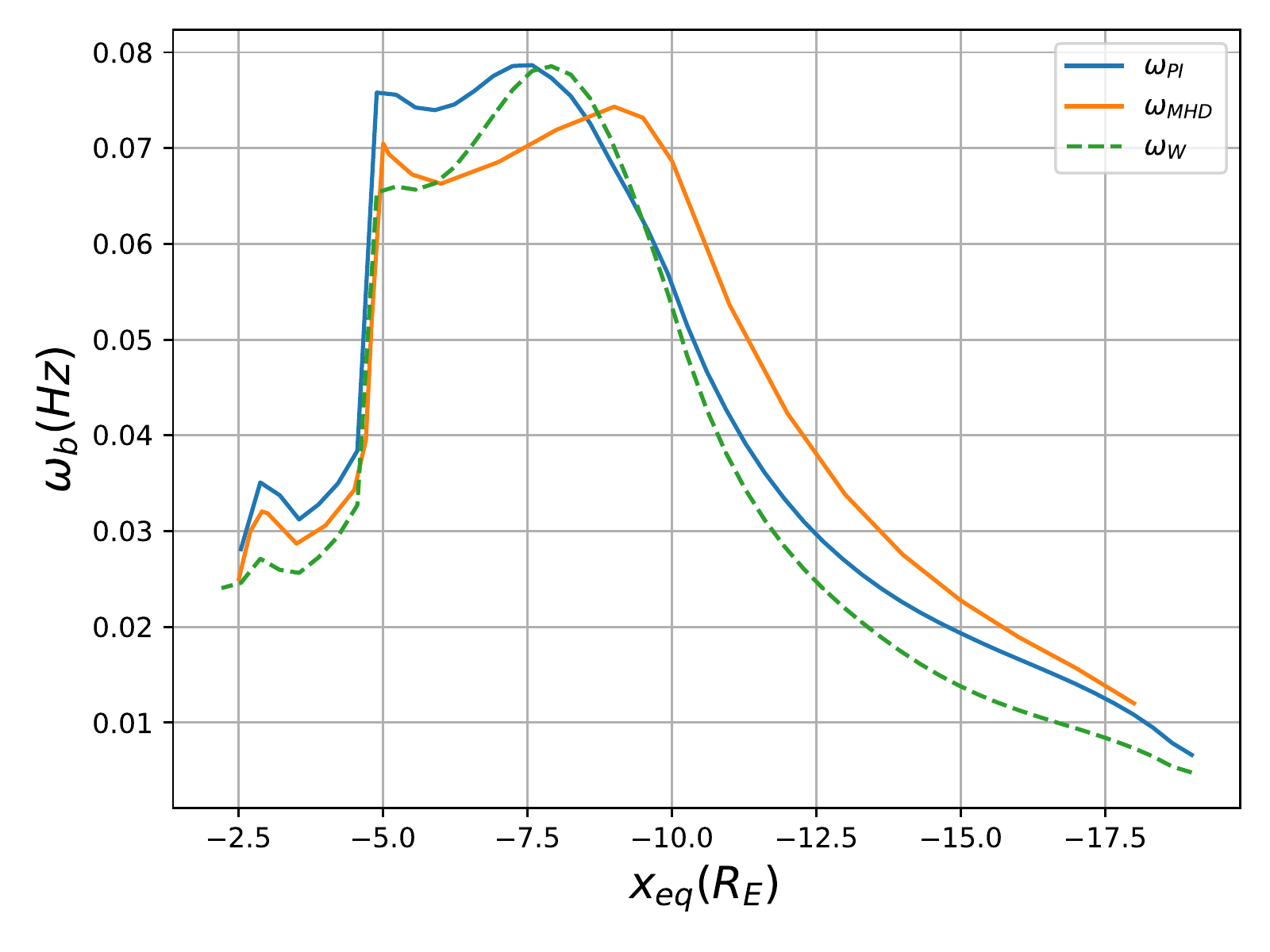}
	\caption{Comparison of eigenfrequencies for pure interchange mode $\omega_{PI}$, MHD normal mode $\omega_b$ (with zero conductivity at the ionosphere), and a simple harmonic oscillator formula for a thin filament with mass concentrated at the equatorial plane $\omega_W$. All three are plotted as a function of $x_{eq}$, the equatorial crossing distance for the footprint of the filament in $R_E$. They agree to within an order of magnitude (more details below). This makes the classic interchange theory, which assumes that $\vec{B} \cdot \nabla p = 0$, a useful simplified model for a more realistic magnetosphere. Note that this result depends on the density model described in point 4 of the preceding section. The sharp gradient in frequency at around $x_{eq} = 5 \ R_E$ is due to the plasmapause.}
	\label{Figure6}
\end{figure}

Figure \ref{Figure6} plots three different eigenfrequencies: (i) the pure interchange oscillation frequency $\omega_{PI}$ derived in equation \ref{39}; (ii) the oscillation frequencies $\omega_{MHD}$ for MHD normal modes of thin filaments in an average magnetosphere (described in Section 3.1); and (iii) the frequency of oscillations of the thin filament treated as a simple harmonic oscillator in a simple plasma sheet magnetic field model with most of its mass concentrated at the equatorial plane (equation (33) of \citeA{WOLF2012} and equation \ref{42} below). The comparison between the first two of these is the most important, as it is one indicator of how much the MHD normal modes resemble those of classic interchange theory, at least for an average magnetosphere.

The main frequency we're comparing to is that of the MHD thin filaments.  In \citeA{TOFFO2020}, numerical solutions to an eigenvalue problem consisting of two coupled differential equations for the normal modes $\xi_\perp(s)$ and $\xi_\parallel(s)$ at different values of $x_{eq}$. In effect, one arrives at a detailed picture for a realistic magnetosphere of the normal modes in the $xz$ plane along with the corresponding buoyancy frequency $\omega_b$ as a function of equatorial crossing point $x_{eq}$. This frequency is determined numerically and has no analytic expression. These normal modes themselves will be discussed in greater detail shortly. The MHD thin filament frequency has been determined numerically from the oscillations of the lowest even mode.

We define relative differences from the MHD normal modes as follows:

\bqy
	\sqrt{ \left< \left( \log_{10}{\left[\frac{\omega_{PI}} {\omega_{MHD}}\right]} \right)^2 \right>_{x_{eq}} } &=& 0.054 \\
	\sqrt{ \left< \left( \log_{10}{\left[\frac{\omega_{W}} {\omega_{MHD}}\right]} \right)^2 \right>_{x_{eq}} } &=& 0.112
\eqy

The first relative difference, again, is the most important for our purposes. The agreement between the frequency from the classic interchange theory and that of the MHD normal mode results is quite good. Note that here our differences contain averages over equatorial crossing point $x_{eq}$, denoted $\left< f \right>_{x_{eq}}$.

\subsection{Eigenfunction Calculation}

We can also examine the normal mode eigenfunctions and compare them against the pure interchange modes for fieldlines that have various equatorial crossing points. The agreement is quite pronounced, especially in the plasma sheet. As one approaches the inner magnetosphere, the deviation between the parallel modes is greater. In the transition region between $x_{eq} = -11 \ R_E$ and $x_{eq} = - 6 \ R_E$, the perpendicular modes also deviate from one another, but the agreement is somewhat closer Earthward of that.

\begin{figure}[h!]
	\centering
	\textbf{Displacement Eigenfunctions for Pure Interchange and MHD Thin Filaments}
	\includegraphics[width=1\textwidth]{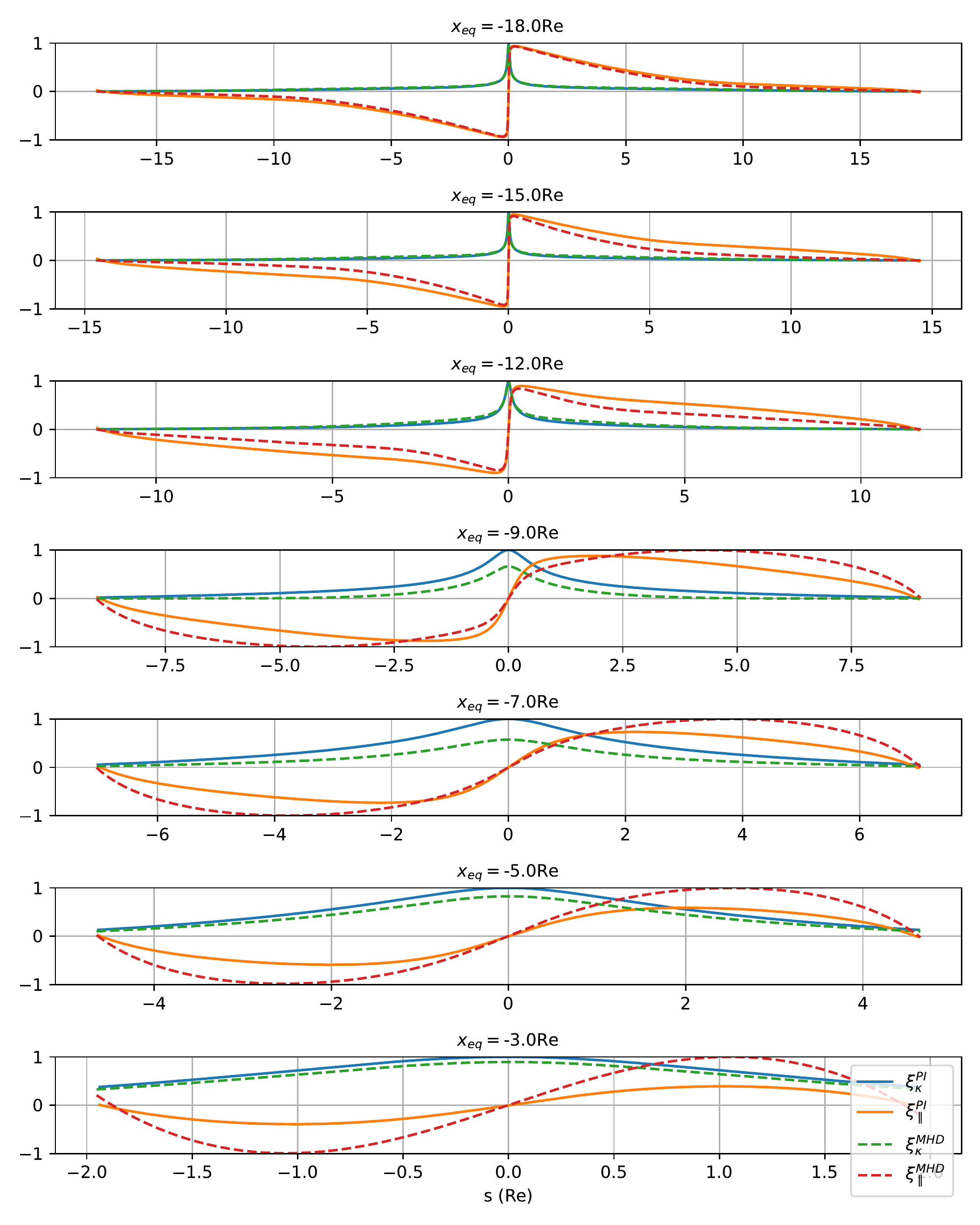}
	\caption{Displacements of mass points parallel and perpendicular to the magnetic field lines at various equatorial crossing points ($x_{eq}$) from the inner magnetosphere to the further out in the plasma sheet. Solid lines show the displacements according to the classic interchange theory, while dashed lines show the results of MHD normal mode calculations. These deviate further from one another as one approaches the inner magnetosphere. Displacements are normalized to their maxima. Recall that $\xi_\kappa = - \xi_\perp$.}
	\label{Figure7}
\end{figure}

Figure \ref{Figure7} shows displacements $\xi_\parallel$ and $\xi_\kappa = - \xi_\perp$ for seven filaments and for pure interchange and MHD full normal modes. We will begin by explaining how the displacements are calculated numerically for pure interchange. Note that these are the displacements normalized to their maxima, denoted in \citeA{TOFFO2020} by $\Xi$. There is no loss of generality since the maxima are arbitrary. Also, note that each of the modes are normalized by the value $max(|\xi_\kappa |, |\xi_\parallel |)$. Note that a positive value of $\xi_\kappa$ represents a fieldline that moves Earthward from its equilibrium point, while a positive/negative value of $\xi_\parallel$ in the northern/southern hemisphere represents a fieldline that gets shorter whereby mass points on the fieldline move closer to the Earth. 

The algorithm for computing the displacements $\xi_\perp$ and $\xi_\parallel$ for the pure interchange and normal modes, plotted in Figure \ref{Figure7} is as follows:
\begin{enumerate}
	\item The first step in this procedure involves determining the field line that crosses the equatorial plane at a specified location $x$ (labeled $F_0$) and another nearby field line (labeled $F_1$) that has equatorial crossing location  $x - \delta x$, where $\delta x$ is chosen to be a small number (typically $0.1 \ R_E$). The background magnetic field is based on the same magnetic field model used in the \citeA{TOFFO2020}, which is a \citeA{TSY1989} with an empirical background pressure that has been relaxed to equilibrium using a 2-D version of the friction technique \cite{LEMON2003}.
	\item For each grid point on field line $F_0$, the intersection point on field line $F_1$ along the $\hat{\kappa}$ is found.
	\item The distance from the grid point on field line $F_0$ to the intersection point on field line $F_1$ is used to determine $\xi_\kappa$.
	\item Equation (\ref{37}) is then used to compute $\xi_\parallel$, by computing $V_p(\psi)$, the flux tube volume $V$ and using the locations of the two field lines $F_0$ and $F_1$ to compute the necessary derivatives.
\end{enumerate}

For the normal mode calculation, the basic procedure is described in \citeA{TOFFO2020}, except that the ionospheric boundary condition is replaced with a zero conductance, to be consistent with the interchange assumption. The details of the boundary condition are described in the Appendix. To accurately reproduce the boundary condition, and satisfy equations (\ref{A4}) and (\ref{A12}) it was necessary to move the location of the inner boundary out to $2 \ R_E$. This is due to the limited number of grid points used to store the background magnetic field, which was a Cartesian grid with a resolution of $0.03 \ R_E$.

In addition to comparing the eigenmodes for the MHD thin filament calculation and those corresponding to a pure interchange assumption, we can examine just how much the pressure perturbations deviate from constancy for the MHD thin filaments. For pure interchange modes, the $\delta P$ would be constant.

\begin{figure}[h!]
	\centering
	\textbf{MHD Thin Filament Pressure Perturbation}
	\includegraphics[width=1\textwidth]{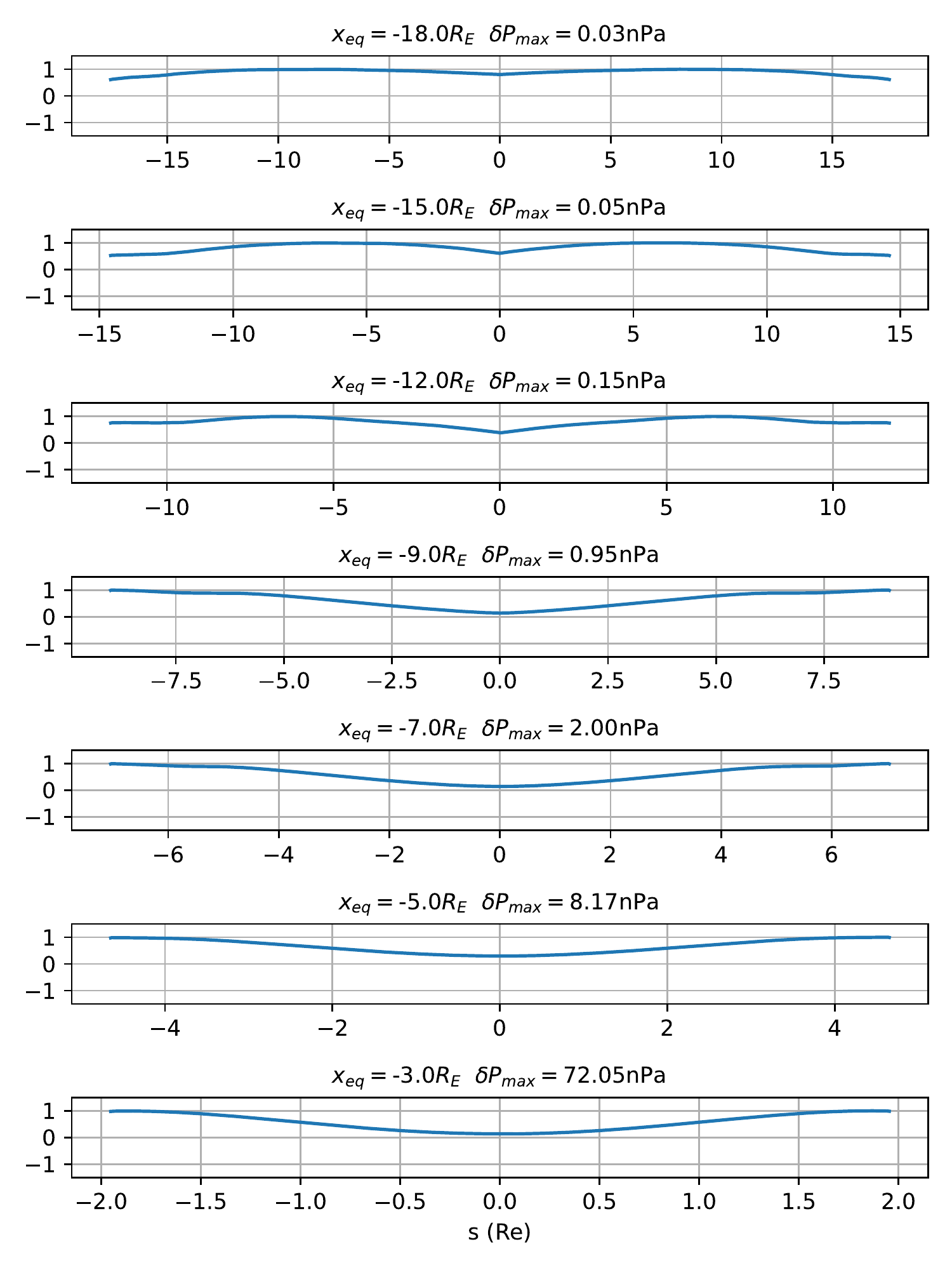}
	\caption{Plot of the pressure perturbation profiles along the field line coordinate $s$ for the MHD thin filaments at various equatorial crossing points $x_{eq}$. They are normalized to the maximum pressure, given above each panel. Note that the pressure perturbation deviates little from constancy out in the plasma sheet, but deviates more substantially as one approaches the inner magnetosphere.}
	\label{Figure8}
\end{figure}

Table \ref{table:table1} shows the root-mean-square differences between the pure interchange and MHD normal modes at each equatorial crossing point value. Note that here our differences involve averages along the field line $s$. This is different than the flux tube average $\left< f \right>_{ft}$ defined above.

\begin{table}
\begin{center}
	\begin{tabular}{ c   c  c }
		\hline
		Equatorial crossing point \\$x_{eq}  (R_E)$ & $\sqrt{ \left< \left( \xi_\kappa^{PI} - \xi_\kappa^{MHD} \right)^2 \right>_s }$ & $\sqrt{ \left< \left( \xi_\parallel^{PI} - \xi_\parallel^{MHD} \right)^2 \right>_s }$ \\ \hline
		 - 3 & $0.0958$ & $0.357$  \\ %\hline
		 - 5 & $0.136$ & $0.252$ \\ %\hline
		 - 7 & $0.297$ & $0.214$ \\ %\hline
	 	 - 9 & $0.199$ & $0.202$ \\ %\hline
		 - 12  & $0.0336$ & $0.151$ \\ %\hline
		 - 15  & $0.0172$ & $0.135$ \\ %\hline
		 - 18  & $0.00983$ & $0.0388$ \\ \hline		
	\end{tabular}
	\caption{Root-mean-square differences, averaged over a fieldline $s$, between the pure interchange (PI) and MHD eigenfunctions for both the $\kappa$ and parallel modes for various fieldlines that cross the equatorial plane at  $x=x_{eq}$.}
	 \label{table:table1}
\end{center}
\end{table}

\section{Discussion and Conclusion}

The most striking feature of the eigenfunctions is how different they are in the plasma sheet and the inner magnetosphere. In the plasma sheet ($x_{eq} = - 12 \ R_E$ to $x_{eq} = - 18 \ R_E$), $\xi_\perp$ is very concentrated at the equatorial plane. $\xi_\parallel = 0$ at the equatorial plane (because of symmetry), but $\left| \xi_\parallel \right|$ has a very sharp maximum just off the equatorial plane. (See Figure \ref{Figure7}.) Earthward of that maximum, $\left| \xi_\parallel \right|$ gradually declines earthward. The total velocity is basically in the $x$-direction. It is also striking how similar the eigenfunctions for interchange and MHD modes are. The interchange and MHD eigenfunctions are a little different near the inner edge of the plasma sheet, but they are still very similar. Note that in the plasma sheet, $\nabla \log{P V^\gamma}$ is tailward but very small (right side of Figure \ref{Figure5}). We associate all of these features with buoyancy waves.

The form of the eigenfunctions are dramatically different for the inner magnetosphere ($x_{eq} = - 3 \ R_E$ to $x_{eq} = - 7 \ R_E$). $\xi_\perp$ and $\xi_\parallel$ are now approximately sinusoidal and do not show strong peaks. These eigenfunctions qualitatively resemble classic slow modes, and we don't refer to them as buoyancy waves. The interchange and full MHD eigenfunctions are not exactly the same, but they are not strikingly different. $\nabla \log{P V^\gamma}$ is tailward and much stronger than in the plasma sheet. The interchange and MHD eigenfrequencies are again very similar.

Figure 8 indicates that $\delta P$ always has a minimum at the equatorial plane. The $\delta P$ varies very modestly for a filament that crosses the equatorial plane at $x_{eq} = - 18 \ R_E$, and that is consistent with the fact that interchange and MHD eigenfrequencies and eigenfunctions are almost the same. $\delta p$ varies more substantially as $\left| x \right|$ decreases.

It is helpful to discuss the Wolf 2012 formula displayed in Figure \ref{Figure6} in \citeA{WOLF2012}, which identifies the oscillation frequency with that of a simple harmonic oscillator undergoing small oscillations about the equatorial crossing point $x_{eq}$ of the filament. It assumes that the effective net force on the filament is just the force difference between force transverse to the filament and in the background. The magnetic field is assumed to have a simple form $\vec{B} = B_{ze} + B_x' z \hat{x}$, where $B_{ze}$ and $B_x'$ are constants. This approach uses Newton's second law for the net force in the $x$-direction at the equator and treats the mass as oscillating with the equatorial crossing point at frequency $\omega_W$. The result is:

\bq
	\omega_W^2(x) = \frac{\pi P}{\rho B_{z,eq} V_{eq}} \frac{1}{\left( 1 + \frac{5 \left< \beta \right>_{ft}}{6}\right)} \frac{K_{eq}'}{K_{eq}} \approx \frac{\left( 0.0741 \, Hz^2 \right)}{\left( 1 + \frac{5 \left< \beta \right>_{ft}}{6}\right)} \frac{T_i}{B_{z,eq} V_{eq} } \frac{K_{eq}'}{K_{eq}}, \label{42}
\eq

where the prime denotes a derivative wrt to $x$ in equatorial crossing point $x_{eq}$. In this calculation, it was assumed that electrons are cold and ions are protons. Figure \ref{Figure6} indicates that the \citeA{WOLF2012} formula is close to interchange and MHD frequencies. Ion temperature has $keV$ units in the second formula. The same formula also is reasonably consistent with ULF wave observations (Figure 4 of \citeA{PANOV2013}). It is remarkable that equation (\ref{42}) was derived the plasma sheet ($\vec{B} = B_{ze} + B_x' z \hat{x}$), but it also works quite well for the inner magnetosphere. It is not clear why that is true, it could be that the determining factor for the eigenfrequencies is the $\nabla \log{P V^\gamma}$; we will address this in a future study.

Both the classic interchange and MHD thin filament treatments made use of simplifying assumptions beyond the standard assumptions implicit in the relevant theories.  Some of these simplifications are shared between the two treatments.  For example, both are one-dimensional, neglecting motion of the flux tubes transverse to the noon-midnight meridional ($xz-$)plane.  However, for the interchange of two flux tubes to occur, they must move past one another in such a way that both cannot remain in-plane.  Both treatments also ignore feedback from the background fields, assuming that the filaments glide freely through the magnetosphere without influence from their surroundings.

However, the difference in simplifying assumptions between the two models is more relevant to their comparison than the oversimplifications they share.  For example, the MHD thin filament code neglects feedback not only from the background, but from any other filament.  This coupling is explicitly included in the classic interchange derivations performed above, hence the need for color indices to reference specific filaments.  Strictly speaking, the coupling to the other filament must be included to properly compare some features of the MHD thin filament model to the classic interchange model.  It is surprising how well the frequencies agree given that the MHD thin filament code neglects this important coupling aspect of the interchange process.

\appendix

\section{Conductivity Model}

In this section, we derive boundary conditions that are suitable for comparing the classic interchange theory with the modes one obtains from MHD calculations for thin filaments. To meaningfully include the interchange of flux tubes, we must use zero ionospheric conductance, as the filaments must be free to move in the ionosphere.

\begin{figure}[h!]
	\centering
	\includegraphics[width=1\textwidth]{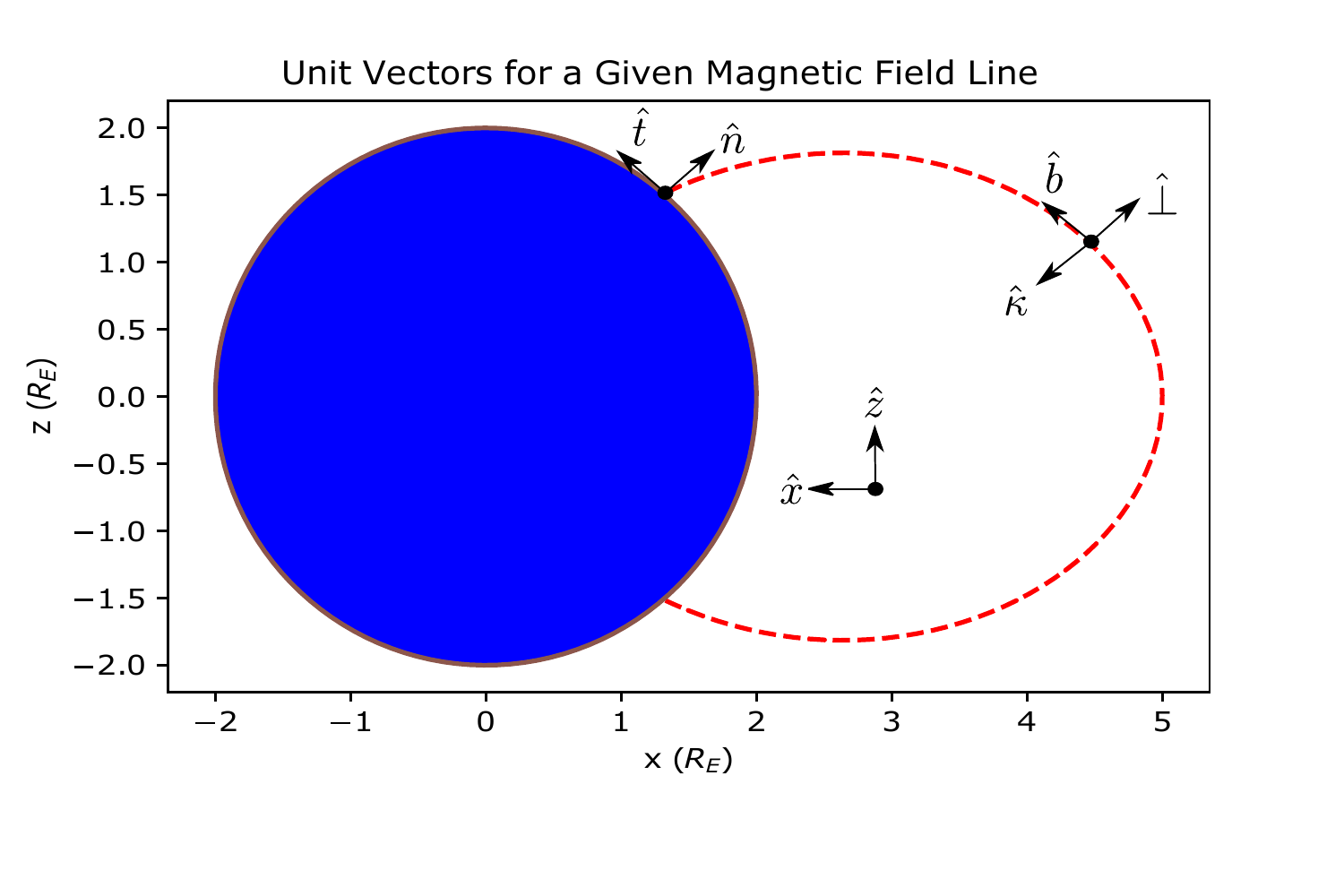}
	\caption{Shown above is an individual magnetic field line, along with all of the relevant unit vectors used in various calculations throughout this paper. There are three coordinate systems: one is the standard fixed Cartesian $xyz$ coordinate system, second is relative to the Earthward boundary footprint of the field line, normal $\hat{n}$ and tangential $\hat{t}$ to that boundary, and the third is local and relative to the field-line directions, with field-transverse $\hat{\kappa}$ and field-aligned $\hat{b}$ unit vectors. Note that $\hat{\perp} = - \hat{\kappa}$.}
	\label{Figure9}
\end{figure}

Let $\hat{n}$ be the outward normal to the modeling region, $\hat{y}$ the dawn-to-dusk direction transverse to the plane of calculation and $\hat{t}$ the northward tangent to the boundary.  Note that we have $\hat{y} = \hat{b} \times \hat{\kappa} = \hat{n} \times \hat{t}$. See Figure \ref{Figure9} above for a reminder of the various coordinate systems and unit vectors involved in our calculations.

Boundary conditions on the velocity:

\bq
	\vec{v} = v_\parallel \hat{b} + \frac{E_y \hat{y} \times \vec{B}}{B^2}
\eq

will be translated into conditions on the linear wave displacements $\xi_\parallel$ and $\xi_\kappa$ using $\vec{v} = - i \omega \vec{\xi}$.

The normal boundary velocity vanishes $\vec{v} \cdot \hat{n} = 0$, giving us a condition on the parallel displacement:

\bq
	\xi_\parallel = \frac{1}{i \omega} \frac{\hat{b} \cdot \hat{t}}{\hat{b} \cdot \hat{n}} \frac{E_y}{B}.
\eq

We can similarly determine the curvature-directed boundary displacement in terms of the fields:

\bq
	\xi_\kappa = - \frac{1}{i \omega} \frac{E_y}{B}.
\eq

This allows us to arrive at one boundary condition relating the two displacements:

\bq
	\xi_\parallel = - \frac{\hat{b} \cdot \hat{t}}{\hat{b} \cdot \hat{n}} \xi_\kappa. \label{A4}
\eq

Our second boundary condition can be obtained from the tangential boundary velocity:

\bq
	\vec{v} \cdot \hat{t} = \frac{E_y}{B_n} \left[ \left( \hat{\kappa} \cdot \hat{t} \right) \left( \hat{b} \cdot \hat{n} \right) - \left( \hat{b} \cdot \hat{t} \right)^2 \right],
\eq

which in general depends on the conductance of the boundary. Note that we have abbreviated $\vec{B} \cdot \hat{n}$ as $B_n$. The current line density along the west side of the filament ($y = \epsilon$) is given by:

\bq
	\vec{j} = - \frac{\hat{y} \times \Delta\vec{B}}{\mu_0} \delta(y - \epsilon),
\eq

where $\Delta\vec{ B} \coloneqq \vec{B}_{filament} - \vec{B}_{background}$. This current is completed across the ionospheric footprint by the Pedersen conduction current, so we have:

\bq
	\Sigma_p E_y = - \frac{\Delta\vec{ B} \cdot \hat{t}}{\mu_0}. \label{A7}
\eq

We write $\Delta\vec{ B} \cdot \hat{t}$ in terms of the linear wave displacements using equations $\left( 21 \right) - \left( 25 \right)$ and $\left( 44 \right) - \left( 45 \right)$ from \citeA{TOFFO2020} and eliminate the electric field from the tangent boundary velocity using (\ref{A7}) to arrive at the following condition:

\bq
	f_s \left( \hat{b} \cdot \hat{t} \right) \left( \p_s \xi_\parallel - \frac{\p_s B}{B} \xi_\parallel - 2 \kappa \xi_\kappa \right) + \left( \hat{\kappa} \cdot \hat{t} \right) \left( \p_s \xi_\kappa - C \xi_\kappa \right) = i \omega \mu_0 \Sigma_p \frac{B_n}{B} \frac{\left[ \left( \hat{b} \cdot \hat{t} \right) \xi_\parallel + \left( \hat{\kappa} \cdot \hat{t} \right) \xi_\kappa \right]}{\left[ \left( \hat{\kappa} \cdot \hat{t} \right) \left( \hat{b} \cdot \hat{n} \right) - \left( \hat{b} \cdot \hat{t} \right)^2 \right]},
\eq

where:

\bqy
	f_s &\coloneqq& \frac{c_s^2}{c_s^2 + c_A^2}, \\
	f_A &\coloneqq& \frac{c_A^2}{c_s^2 + c_A^2}, \\
	C &\coloneqq& \hat{\kappa} \cdot \left[ \left( \hat{\kappa} \cdot \nabla \right) \hat{b} \right].
\eqy

This is a general condition for thin filament calculations that work for boundary orientation and ionospheric conductance. However, as mentioned above, compatibility with the classic interchange theory demands that the ionospheric conductance vanish.  Assuming zero conductance and $f_s \ll 1$, we arrive at our second boundary condition:

\bq
	\p_s \xi_\kappa = C \xi_\kappa, \label{A12}
\eq

where $C$ has been calculated numerically. We can see that the zero conductance case has mixed (Neumann and Dirichlet) boundary conditions. The reason that $\p_s \xi_\kappa$ must be a nonvanishing fraction of $\xi_\kappa$ is that the direction of the background field lines depends on position so that the filament line must rotate when displaced in order to remain parallel to the local background.

\acknowledgments
This work is supported by NASA Heliophysics supporting research grant 80NSSC18K1226 and NASA theory and modeling grant 80NSSC20K1276.

\bibliography{PI}

\end{document}